# A simple algorithm for decoding Reed-Solomon codes and its relation to the Welch-Berlekamp algorithm

Sergei Fedorenko

**Abstract**

A simple and natural Gao algorithm for decoding algebraic codes is described. Its relation to the Welch-Berlekamp and Euclidean algorithms is given.

**Index Terms**

decoding algorithm, decoding Reed-Solomon codes, Euclidean algorithm, key equation, Reed-Solomon codes, remainder decoding, Welch-Berlekamp algorithm.

## I. INTRODUCTION

In the recent article Gao [1] described a simple and natural algorithm for decoding algebraic codes in the class of algorithms decoding up to the designed error correcting capability. The asymptotic complexity of this algorithm coincides with the complexity of the best Reed-Solomon decoding algorithms, and the description is the simplest for known algorithms descriptions. In this paper the Gao algorithm's relation to the Welch-Berlekamp [2] and Euclidean algorithms [3], [4] is given.

## II. DEFINITIONS AND NOTATIONS

Let us define the $(n, k, d)$ Reed-Solomon (RS) code over GF$(q)$ with length $n = q - 1$, number of information symbols $k$, designed distance $d = n - k + 1$, $q$ is prime power.

The RS code generator polynomial is

$$g(x) = \prod_{i=b}^{b+d-2} (x - \alpha^i),$$

where $\alpha$ is a primitive element of GF($q$), $b$ is any natural number.

The received vector is represented as a polynomial

$$R(x) = \sum_{i=0}^{n-1} r_i x^i = C(x) + E(x) = \sum_{i=0}^{n-1} c_i x^i + \sum_{i=0}^{n-1} e_i x^i,$$

where $C(x)$ is the codeword, $E(x)$ is the error vector.

The $i$th error in the error vector $E(x)$ has a locator $Z_i \in \{\alpha^0, \alpha^1, \alpha^2, \cdots, \alpha^{n-1}\}$ and an error value $Y_i \in$ GF($q$)\0.

The error locator polynomial is

$$W(x) = \prod_{i=1}^{t} (x - Z_i),$$

where $t \leq (d-1)/2$, $Z_i$ is the locator of the error position in the error vector $E(x)$.

There are several coding methods for RS codes. In this paper two methods are used: the remainder coding in the time domain for systematic coding and the spectral coding in the frequency domain for non-systematic coding. Note that the decoding algorithm does not depend on the coding method.

In Sections 4 and 5 remainder coding is applied. The codeword consists of two parts: the message part and the parity part. The message error locator polynomial is

$$W_m(x) = \prod_{i=1}^{t_m} (x - Z_i),$$

where $t_m \leq (d-1)/2$, $Z_i \in \alpha^j$, $j \in [d-1, n-1]$, is the locator of the error position in the message part of error vector $E(x)$.

In Sections 3 and 6 spectral coding is applied. The message polynomial of RS code is

$$M(x) = \sum_{i=0}^{k-1} m_i x^i.$$

The component $c_i$ of codeword $C(x)$ is computed as

$$c_i = M(\alpha^i), \quad i \in [0, n-1].$$

## III. GAO ALGORITHM

We describe here the Gao algorithm [1] for the case of the classical RS codes only ($n = q-1$).

1) Interpolation.

    Construct an interpolating polynomial $T(x)$ such that

    $$T(\alpha^i) = r_i, \quad i \in [0, n-1],$$

    where $\deg T(x) < n$.

2) Partial GCD.

    Solve a congruence

    $$\begin{cases} W(x)T(x) \equiv P(x) \bmod x^n - 1 \\ \deg P(x) < \frac{n+k}{2} \\ \text{maximize } \deg P(x) \end{cases}$$

    by applying the extended Euclidean algorithm to $x^n - 1$ and $T(x)$, and we obtain unique pair of polynomials $P(x)$ and $W(x)$.

3) Division.

    The message polynomial is

    $$M(x) = \frac{P(x)}{W(x)}.$$

The asymptotic complexity of this algorithm $O(n(\log n)^2)$ coincides with the complexity of good classical RS decoding algorithms [4], [5].

The first step of the Gao algorithm can be implemented by any fast algorithm for the discrete Fourier transform, for example, [6]. If the number of multiplications is to be minimized, then the best known algorithm of the discrete Fourier transform for small lengths ($n \leq 511$) is described in [7].

The best implementation of the second step is the Moenck algorithm [8], that is also reproduced in [9], [10]. Note that Moenck's implementation of the second step completely coincides with the algorithm of the key equation solving by Sugiyama *et al.* [3].

For application of the Gao algorithm to other classes of algebraic codes, such as BCH, Goppa, or alternant codes, the additional re-encoding step is needed.

## IV. THE ORIGINAL WELCH-BERLEKAMP ALGORITHM

The Welch-Berlekamp algorithm is the remainder decoding algorithm. We follow the original patent [2] and paper [11] with simple algorithm description.

The Welch-Berlekamp algorithm consists of the following steps:

1) Calculate the syndrome for cyclic codes

$$S(x) = \sum_{i=0}^{d-2} s_i x^i \equiv R(x) \bmod g(x).$$

2) Solve the key equation

$$\begin{cases} p_j \alpha^j N(\alpha^j) = s_j W_m(\alpha^j), & j \in [0, d-2] \\ \deg N(x) < \deg W_m(x) \leq (d-1)/2, \end{cases} \quad (1)$$

where

$$p(x) = \frac{g(x)}{x - \alpha^b} = \sum_{i=0}^{d-2} p_i x^i,$$

and obtain polynomials $N(x)$ and $W_m(x)$.

3) Determine the error locations in the message part of received vector from $W_m(x)$ and obtain the set $\{Z_i\} \in \alpha^j$, $j \in [d-1, n-1]$.

4) Determine the error values in the message part of received vector

$$Y_i = f(Z_i) \frac{N(Z_i)}{W'_m(Z_i)},$$

where $W'_m(x)$ is a formal derivative for $W_m(x)$,

$$f(Z) = Z^{-b} \sum_{i=0}^{d-2} \frac{p_i \alpha^{i(b+1)}}{\alpha^i - Z}$$

for $Z \in \alpha^j$, $j \in [d-1, n-1]$.

5) Determine the parity part of codeword.

## V. CHAMBERS INTERPRETATION

In [12] Chambers described the second step interpretation of the Welch-Berlekamp algorithm.

For RS codes $p_j \neq 0$, $j \in [0, d-2]$, and from the key equation (1) we have

$$\begin{cases} N(\alpha^j) = \frac{s_j}{p_j \alpha^j} W_m(\alpha^j), & j \in [0, d-2] \\ \deg N(x) < \deg W_m(x) \leq (d-1)/2. \end{cases}$$

Construct an interpolating polynomial $L(x)$ such that

$$L(\alpha^j) = \frac{s_j}{p_j \alpha^j}, \quad j \in [0, d-2],$$

where $\deg L(x) < d - 1$.

From

$$N(\alpha^j) = L(\alpha^j) W_m(\alpha^j), \quad j \in [0, d-2],$$

we have

$$\begin{cases} N(x) \equiv L(x) W_m(x) \bmod g(x) \\ \deg N(x) < \deg W_m(x) \leq (d-1)/2 \\ \text{minimize } \deg W_m(x). \end{cases}$$

Solve this congruence by applying the extended Euclidean algorithm to $g(x)$ and $L(x)$, and we obtain polynomials $N(x)$ and $W_m(x)$.

## VI. ANOTHER APPROACH TO THE WELCH-BERLEKAMP ALGORITHM

In [13] Gemmell and Sudan described a modification of the Welch-Berlekamp algorithm which is easy to understand. However, direct implementation of this modification is inefficient.

If $r_i = c_i$ then $r_i = M(\alpha^i)$. If $r_i \neq c_i$ then $W(\alpha^i) = 0$. Hence,

$$W(\alpha^i) r_i = W(\alpha^i) M(\alpha^i), \quad i \in [0, n-1].$$

Let $P(x) = W(x) M(x)$. Then the key equation is

$$W(\alpha^i) r_i = P(\alpha^i), \quad i \in [0, n-1]. \tag{2}$$

We solve the key equation and obtain polynomials $P(x)$ and $W(x)$. The message polynomial is

$$M(x) = \frac{P(x)}{W(x)}.$$

The Gemmell-Sudan modification applies the idea of the original Welch-Berlekamp algorithm to the frequency domain. Clearly, the description of the Gemmell-Sudan modification is simpler than the description of the original Welch-Berlekamp algorithm.

## VII. Improvement of the Gemmell-Sudan modification

We apply the Chambers method to the key equation (2).

Let us construct an interpolating polynomial $T(x)$ such that

$$T(\alpha^i) = r_i, \quad i \in [0, n-1],$$

where $\deg T(x) < n$. This corresponds to the first step of the Gao algorithm.

Further, from

$$W(\alpha^i)T(\alpha^i) = P(\alpha^i), \quad i \in [0, n-1],$$

we have the congruence

$$W(x)T(x) \equiv P(x) \bmod x^n - 1.$$

Solving this congruence corresponds to the second step of the Gao algorithm.

It is clear that by combining methods from Sections 5 and 6, we get the Gao algorithm.

## VIII. Conclusion

The paper does not contain any new decoding algorithms. However, the demonstrative connection between known decoding algorithms has important methodological significance in the coding theory. We see that the Gao algorithm belongs to the class of the frequency domain algorithms with simple description. Thus the paper demonstrates how carrying the algorithms from the time domain into the frequency domain results in simplification of their description. In author's opinion the Gao algorithm and its modifications are the simplest for codes with short lengths for any implementation. The correctness proofs of all considered algorithms are simple and can be found in corresponding references. The properties of the extended Euclidean algorithm are described in popular books on the coding theory [4], [14].

**Sergei Valentinovich Fedorenko** (Member, IEEE'01) was born in St.Petersburg, U.S.S.R. in 1967. He received the Ph.D. degree in computer science from St.Petersburg State University of Airspace Instrumentation in 1994. Currently he is an Associate Professor in the Distributed Computing and Networking Department, St.Petersburg State Polytechnic University (194021, Polytechnicheskaya str., 21, St.Petersburg, Russia. e-mail: sfedorenko at ieee dot org). His research interests include error-correcting codes, decoding algorithms, discrete Fourier transform over finite fields, and communication systems.